\documentclass[aps,prl,preprint,groupedaddress]{revtex4-1}
\usepackage{theorem,amsmath,amssymb}
\usepackage{enumerate}
\newcommand{\cE}{{\mathord{\cal E}}}
\newcommand{\x}{{\mathbf{r}}}
\renewcommand{\j}{\mathbf{j}}
\newcommand{\A}{\mathbf{A}}
\newcommand{\aA}{\mathbf{a}}
\newcommand{\eps}{\varepsilon}
\newcommand{\cA}{{\mathord{\cal A}}}

\begin{document}


\title{Non-existence of a Hohenberg-Kohn Variational Principle in Total Current Density Functional Theory}


\author{Andre Laestadius and Michael Benedicks}
\email[]{andrela@math.kth.se and michaelb@math.kth.se}
\affiliation{Department of Mathematics, KTH Royal Institute of Technology, Sweden}


\date{\today}

\begin{abstract}
For a many-electron system, whether the particle density $\rho(\x)$ 
and the total current density $\j(\x)$ are sufficient 
to determine the one-body potential $V(\x)$ and vector potential $\A(\x)$, is still an open question. 
For the one-electron case, a Hohenberg-Kohn theorem exists formulated with the total current density. Here we show that 
the generalized Hohenberg-Kohn energy functional $\cE_{V_0,\A_0}(\rho,\j) = \langle \psi(\rho,\j),H(V_0,\A_0)\psi(\rho,\j)\rangle$ can be 
minimal for densities that are not the ground-state densities of the fixed potentials $V_0$ and $\A_0$. Furthermore, for an arbitrary number of electrons and 
under the assumption that a Hohenberg-Kohn theorem exists formulated with $\rho$ and $\j$, we show that 
a variational principle for Total Current Density Functional Theory 
as that of Hohenberg-Kohn for Density Functional Theory does not exist. The reason is that the assumed map 
from densities to the vector potential, written $(\rho,\j)\mapsto \A(\rho,\j;\x)$, enters explicitly in $\cE_{V_0,\A_0}(\rho,\j)$.
\end{abstract}

\pacs{}

\maketitle

\section{I. INTRODUCTION}
A cornerstone of modern density functional theory is the Hohenberg-Kohn theorem, which states that knowledge of the one-body density 
of a quantum-mechanical system determines the one-body potential $V$ of the same system \cite{HK64} (see also \cite{Lieb83} 
for a complete 
proof). It has been argued \cite{Sahni2010,Diener} that one can extend this theorem 
to a statement that knowledge of the one-body particle density plus knowledge of the one-body total current density suffices to determine 
the one-body potential and the magnetic vector potential for a many electron system (up to a gauge transformation). Since the proofs in \cite{Sahni2010} and \cite{Diener} 
have been shown to contain errors (cf. \cite{Tellgren,AndreMichael}), the existence of such a theorem 
remains an open question except for the one-electron case. 
For a system with only one electron, however, it can be shown that 
the particle density and the total current density determine the potentials up to a gauge transformation \cite{AndreMichael,Tellgren}. 
Moreover, it is well-known that the same statement made with the paramagnetic current density in place of the 
total current density is not true. 
For instance, Vignale and Capelle \cite{Vignale2002} have constructed a 
counterexample that shows that different Hamiltonians can share a common ground-state 
for systems with magnetic fields. 

In the classical Hohenberg-Kohn theory, one studies a system of $N$ interacting electrons subjected to an electric field. Define the 
system's Hamiltonians to be (in suitable units)
\[H(V) = \sum_{j=1}^N (-\Delta_j + V(\x_j)) + \sum_{j<k}|\x_j-\x_k|^{-1}.\] 
The Hohenberg-Kohn theorem then states that the particle density $\rho(\x)$ determines the potential 
$V(\x)$ up to a constant, and thereby the ground-state of $H(V)$. Let $\cA_N'$ be the set of particle densities having the following property: 
there exists a $V(\x)$ such that $H(V)\psi=e\psi$, $\psi=\psi(\rho)$ is the unique ground-state of $H(V)$ and 
$\rho(\x) = N\int |\psi|^2d\x_2\cdots d\x_N$. 

Now, fix $V_0(\x)$ such that $H(V_0)\psi_0=e_0\psi_0$, 
where $\psi_0$ is the unique ground-state and $\rho_0(\x)$ the ground-state particle density. 
On $\cA_N'$, we can define 
\begin{align*}
\cE_{V_0}'(\rho) =  \langle \psi(\rho),H(V_0)\psi(\rho)\rangle &= \langle \psi(\rho),H(0)\psi(\rho)\rangle + \int \rho(\x) V_0(\x)\\
&= F_{HK}'(\rho)+ \int \rho(\x) V_0(\x),
\end{align*}
where the last equality defines the Hohenberg-Kohn functional $F_{HK}'$. The Hohenberg-Kohn variational principle for Density Functional Theory then states that 
\[e_0 = \cE_{V_0}'(\rho_0) = \min_{\cA_N'}\{ \cE_{V_0}'(\rho)\}. \] 
The map $\rho(\x)\mapsto V(\x)$, guaranteed to exist by the Hohenberg-Kohn theorem, 
does not enter explicitly in the functional to be minimized. 
Also note that $\rho_0$ is the unique minimizer of $F_{HK}'(\rho) + \int \rho V_0$. (To see this, assume that
$\rho'\in \cA_N'$ is another minimizer. Then there is a $V'(\x)\neq V_0(\x) + \text{constant}$, such that $H(V')$ has a 
ground-state $\psi'$ with particle density $\rho'$. Since $V'(\x)\neq V_0(\x) + \text{constant}$, we can conclude that 
$\psi'$ is not the ground-state of $H(V_0)$. But then, $e_0< \langle\psi',H(V_0)\psi' \rangle = \cE_{V_0}'(\rho') =e_0$,
which is a contradiction.)

In this paper, we will discuss the corresponding variational principle 
for Current Density Functional Theory formulated with the 
total current density, i.e., a variational principle for an energy functional $\cE_{V_0,\A_0}$ given by 
\[\cE_{V_0,\A_0}(\rho,\j) = \langle \psi(\rho,\j),H(V_0,\A_0)\psi(\rho,\j)\rangle,\]
where $V_0$ and $\A_0$ are the fixed scalar and vector potential, respectively. 
We will note the following: 
\begin{itemize}
 \item For the energy functional suggested in \cite{Sahni2010}, 
\begin{align*}
 \tilde{\cE}_{V_0,\A_0}(\rho,\j) &=F_{HK}(\rho,\j) + 2\int \j(\x)\cdot \A_0(\x) d\x + \int \rho(\x)(V_0(\x) - |A_0(\x)|^2)d\x \\
 &\neq\cE_{V_0,\A_0}(\rho,\j),
 \end{align*}
the counterexample in \cite{Vignale2002} shows that the densities $\rho(\x)$ and $\j(\x)$ cannot satisfy a variational 
principle \cite{VignaleResponse2013}. We shall here give a mathematical proof of this claim.
  \item For $N=1$ and fixed potentials $V_0$ and $\A_0$, the energy functional $\cE_{V_0,\A_0}(\rho,\j)$ is well-defined since 
  $\rho(\x)$ and $\j(\x)$ determine $V_0(\x)$ and $\A_0(\x)$ up to a gauge transformation \cite{AndreMichael,Tellgren}. However, $\cE_{V_0,\A_0}(\rho,\j)$ 
  can be minimized by other densities then the ground-state densities of $H(V_0,\A_0)$.
  \item For any $N$ and under the assumption that a Hohenberg-Kohn theorem exists formulated with $\rho(\x)$ and $\j(\x)$, we show that 
  no variational principle as that of Hohenberg-Kohn for Density Functional Theory exists for 
  Current Density Functional Theory formulated with the total current density. This is due to the fact that the map $(\rho,\j)\mapsto \A(\rho,\j;\x)$ 
  enters explicitly in $\cE_{V_0,\A_0}(\rho,\j)$. Furthermore, this implies that 
  $\cE_{V_0,\A_0}(\rho,\j)$ cannot be extended to $N$-representable density pairs $(\rho,\j)$ by a Levy-Lieb approach (see Section III B below).
\end{itemize}

\section{II. NON-EXISTENCE OF A VARIATIONAL PRINCIPLE FOR $\tilde{\cE}_{V,\A}$}
Let $N=1$ and consider the Schr\"odinger operator 
$H(V, \A) =  (-i\nabla +  \A)^2 + V(\x)$. 
For simplicity, we will assume that the ground-state is non-degenerate. Let $H_0$ denote 
the Schr\"odinger operator, when the potentials are set to zero, i.e., $H_0 =-\Delta$. 
For the non-degenerate ground-state $\psi_0$, we compute the ground-state particle density and paramagnetic current density from 
$\rho(\x) = |\psi(\x)|^2$ and $\j_p(\x) =\textrm{Im}\,(\psi(\x)^*\nabla\psi(\x))$, respectively. The total current density is then given by the sum 
$\j(\x)= \j_p(\x) + \rho(\x) \A(\x)$.

For $N=1$, a Hohenberg-Kohn theorem exists formulated 
with the total current density \cite{AndreMichael,Tellgren}, i.e., $\rho(\x)$ and $\j(\x)$ determine $V(\x)$ and $\A(\x)$ 
up to a gauge transformation. 
In particular, $\A(\x) = \aA(\rho,\j;\x) - \nabla \chi(\x)$ for some function $\chi(\x)$. Denote by $\cA_1$ the set of density pairs 
$(\rho,\j)$ such that a ground-state 
$\psi_0=\psi(\rho,\j)e^{i\chi(\x) }$ exists and 
fulfills

\begin{itemize}
\item[(i)] $|\psi_0|^2=\rho(\x)$,
\item[(ii)] $\textrm{Im}\,(\psi_0^*\nabla\psi_0)+|\psi_0|^2 \A(\x) = \textrm{Im}\,(\psi^*(\rho,\j)\nabla\psi(\rho,\j)) +|\psi(\rho,\j)|^2\aA(\x) = \j(\x)$,
\item[(iii)] $ H(V,\A)\psi_0 =e_0\psi_0$.
\end{itemize} 
On $\cA_1$, we can define the generalized Hohenberg-Kohn 
functional 
\begin{align}
F_{HK}(\rho,\j)= \langle \psi(\rho,\j),H_0\psi(\rho,\j)\rangle.
\label{HKfunc}
\end{align}
Furthermore, for fixed $V_0(\x)$ and $\A_0(\x)$, we 
define on $\cA_1$, as in \cite{Sahni2010}, the energy functional
\begin{align*}
\tilde{\cE}_{V_0,\A_0}(\rho,\j) =F_{HK}(\rho,\j) + 2\int \j(\x)\cdot \A_0(\x) d\x + \int \rho(\x)(V_0(\x) - |A_0(\x)|^2)d\x.
\end{align*}
Note that $\tilde{\cE}_{V_0,\A_0}(\rho,\j) \neq \langle  \psi(\rho,\j),H(V_0,\A_0)\psi(\rho,\j) \rangle$ on $\cA_1$. Consequently, 
no variational principle for the density pair $(\rho,\j)$ is immediately inherited from the variational principle for 
the wavefunction.

The following facts follow from Theorem 4 in \cite{AndreMichael}: 
\begin{itemize}
 \item For $B<0$ small enough and $0<|\tilde{B}|<|B|$, there exists a wavefunction $\psi_0$ that is the ground-state of both $H(V,\A)$ and 
 $H(\tilde{V},\tilde{\A})$, where $\A(\x)= (B/2)\,\hat{e}_z\times \x$ and 
 $\tilde{\A}(\x)= (\tilde{B}/2)\,\hat{e}_z\times \x$.
 \item With $\eps= (\tilde{B}-B)/2>0$ and $\j_\eps(\x) =\j_0(\x) + \eps(\rho_0\, \hat{e}_z\times \x)$, where $\rho_0$ and $\j_0$ are the 
 ground-state densities computed from $\psi_0$, it follows that $F_{HK}(\rho_0,\j_0)= F_{HK}(\rho_0,\j_\eps)$ for $\eps>0$ sufficiently small.
 \item The density pair $(\rho_0,\j_\eps)$ belongs to $\cA_1$ for $\eps>0$ sufficiently small.
\end{itemize}
 
We now note that 
\begin{align*}
\tilde{\cE}_{V,\A}(\rho_0,\j_\eps) &= F_{HK}(\rho_0,\j_\eps) + 2\int \j_\eps(\x)\cdot \A(\x)d\x + \int\rho_0(\x)(V(\x)- |\A(\x)|^2)d\x\\
&= \tilde{\cE}_{V,\A}(\rho_0,\j_0) + \eps B \int\rho_0(\x) (\hat{e}_z\times \x)^2d\x,
\end{align*}
by the above facts. Consequently, $$\tilde{\cE}_{V,\A}(\rho_0,\j_\eps) - \eps B\int \rho_0(\x)(x^2 + y^2)d\x = \tilde{\cE}_{V,\A}(\rho_0,\j_0),$$ which 
implies $\tilde{\cE}_{V,\A}(\rho_0,\j_\eps) < \tilde{\cE}_{V,\A}(\rho_0,\j_0)$. Thus $\tilde{\cE}_{V,\A}$ does not satisfy a 
variational principle.


\section{III. NON-EXISTENCE OF A HOHENBERG-KOHN VARIATIONAL PRINCIPLE FOR TOTAL CURRENT DENSITY}
\subsection{A. The one-electron case}
Fix $V_0(\x)$ and $\A_0(\x)$ such that 
$H(V_0,\A_0)\psi_0 =e_0\psi_0$, and let $\rho_0(\x)$ and $\j_0(\x)$ denote the ground-state densities. 
On $\cA_1$, we define the energy functional
\begin{align}
\cE_{V_0,\A_0}(\rho,\j) &=F_{HK}(\rho,\j) + 2\int \j(\x)\cdot \A_0(\x) d\x + \int \rho(\x)(V_0(\x) - |\A_0(\x)|^2)d\x \nonumber\\
&-2\int \rho(\x)\A_0(\x)\cdot [\aA(\rho,\j;\x) -\A_0(\x)] d\x,
\label{refunc}
\end{align}
where $[\aA - \A] = 0$ if $\aA - \A =\nabla\chi$ for some $\chi$, and  $[\aA -\A] = \aA - \A$ otherwise. 
From (ii), \eqref{refunc} and the definition of $F_{HK}(\rho,\j)$, it follows that 
\[  \cE_{V,\A}(\rho,\j) = \langle \psi(\rho,\j),H(V,\A)\psi(\rho,\j)  \rangle,  \]
for $(\rho,\j)\in \cA_1$. Furthermore, note that the map $(\rho,\j)\mapsto \aA(\rho,\j;\x)$ enters explicitly in the expression for the functional 
$\cE_{V,\A}(\rho,\j)$.

By the variational principle for the wavefunction, with $\rho(\x)$ and $\j(\x)$ in $\cA_1$, we have
\begin{align*}
\cE_{V_0,\A_0}(\rho,\j) &= \langle \psi(\rho,\j),H(V_0,\A_0)\psi(\rho,\j)  \rangle \\&\geq 
\langle \psi(\rho_0,\j_0),H(V_0,\A_0)\psi(\rho_0,\j_0)  \rangle =\cE_{V_0,\A_0}(\rho_0,\j_0)=e_0.
\end{align*}
Thus $e_0 = \cE_{V_0,\A_0}(\rho_0,\j_0) = \min_{\cA_1} \{ \cE_{V_0,\A_0}(\rho,\j)\}$. 
However, from the facts in Section II, it follows that the minimum of $\cE_{V_0,\A_0}(\rho,\j)$ is not only achieved by the ground-state densities 
$\rho_0(\x)$ and $\j_0(\x)$, but also achieved by infinitely many density pairs $(\rho_0,\j_\eps)$, 
\begin{align}
e_0 = \cE_{V_0,\A_0}(\rho_0,\j_\eps) = \min_{\cA_1} \{ \cE_{V_0,\A_0}(\rho,\j)\}.
\label{remin}
\end{align} Thus, although a variational principle exists for $\cE_{V_0,\A_0}(\rho,\j)$, there is no way of 
knowing whether a minimizer $(\rho,\j)$ also is the ground-state densities of $H(V_0,\A_0)$.\\

\subsection{B. Arbitrary number of electrons}
Now, let $H(V,\A)$ be the Hamiltonian of a system of $N$ electrons and where $N$ is arbitrary, i.e., 
\[H(V,\A) = \sum_{j=1}^N [(-i\nabla_j + A(\x_j))^2 + V(\x_j)]  + \sum_{j<k}|\x_j-\x_k|^{-1}.\]
Under the assumption that a Hohenberg-Kohn theorem could be proven formulated with $\rho$ and $\j$, $\cE_{V,\A}(\rho,\j)$ can  
be defined as in \eqref{refunc} for densities in $\cA_N$. Now $F_{HK}(\rho,\j)$ takes the form
\begin{align*}
F_{HK}(\rho,\j) &= \langle \psi(\rho,\j),(T+W)\psi(\rho,\j) \rangle, \\
T+W &= -\sum_{j=1}^N \Delta_j + \sum_{j<k}|\x_j-\x_k|^{-1},
\end{align*}
which agrees with \eqref{HKfunc} if $N=1$. Here $\cA_N$ is the obvious generalization of $\cA_1$. As in the case 
$N=1$, for $(\rho,\j)\in \cA_N$, we have 
\[ \cE_{V,\A}(\rho,\j) = \langle \psi(\rho,\j),H(V,\A)\psi(\rho,\j)  \rangle. \]
Also, recall that the map $(\rho,\j)\mapsto \aA(\rho,\j;\x)$, which by assumption exists for any $N$, enters explicitly in the expression for 
$\cE_{V,\A}(\rho,\j)$, while in the classical Hohenberg-Kohn theory, the corresponding map $\rho\mapsto V$ 
does not appear in the functional $\cE_{V}'(\rho)$ (see Section I). 
The presence of $\aA(\rho,\j;\x)$ adds an additional layer of complexity to the generalized Hohenberg-Kohn 
energy functional $\cE_{V,\A}$. Furthermore, in the work of Lieb \cite{Lieb83}, $F_{HK}'(\rho)$ is extended to the so called 
Levy-Lieb functional $F_{LL}(\rho) = \inf_{\psi}\{ \langle \psi,H(0) \psi \rangle: \psi\mapsto \rho \}$, 
which is defined on the set of $N$-representable particle densities, denoted $I_N$ (see \cite{Lieb83}). This extension allows 
the functional $F_{LL}(\rho) + \int \rho V$ to be minimized freely on the known set $I_N$ 
instead of the unknown set $\cA_N'$. However, even if $F_{HK}(\rho,\j)$ could be extended to a Levy-Lieb-type functional 
defined for $N$-representable density pairs $(\rho,\j)$, no such extension is possible for $\cE_{V,\A}(\rho,\j)$ because of the term 
$\int \rho \A\cdot (\aA(\rho,\j;\x)-\A)$, which is by definition only meaningful for $(\rho,\j)\in \cA_N$.

\section{ACKNOWLEDGEMENTS}
The authors are very thankful to Fabian Portmann for useful comments and fruitful discussions.

\bibliography{CEreferences}
\end{document}